\documentclass[a4paper,10pt]{article}
\usepackage[dvips]{graphicx}
\begin{document}
{\LARGE Three dimensional  relativistic hydrodynamical model for QGP gas}

\vspace{0.3cm}

\begin{center}
{\Large C.~Nonaka\footnote[1]
{E-mail:nonaka@butsuri.sci.hiroshima-u.ac.jp}, 
S.~Muroya$^*$\footnote[2]{E-mail:muroya@yukawa.kyoto-u.ac.jp} and  
O.~Miyamura\footnote[3]{E-mail:miyamura@fusion.aci.hiroshima-u.ac.jp}}
\end{center}
\begin{center}
{\large \it Dep. of Physics, Hiroshima Univ., Higashi-hiroshima,
Hiroshima, 739, Japan}

{\large \it Tokuyama Women's College, Tokuyama, Yamaguchi, 745, Japan$^*$}
\end{center}
\vspace{0.2cm}
\abstract{We numerically 
solve fully (3+1)-dimensional relativistic 
hydrodynamical equation coupled with the baryon number 
conservation law without spatial symmetry.
We discuss the effect of transverse expansion based on the deviation 
our numerical 
result from 
Bjorken's scaling solution. We analyze the space-time evolution of 
the QGP gas in the case of non cylindrical initial conditions.}

\section{Introduction}
The various kinds of collective flow phenomena such as directed flow, 
elliptic flow 
and radial flow has been observed  
in recent experiments at AGS \cite{flow-AGS} and SPS \cite{flow-SPS}. 
It is a matter of interest that such flow are results of 
hydrodynamical motion of hadronic fluid.
Our first trial to tackle  the problem is to develop (3+1)-dimensional 
hydrodynamical model.
Assuming the local thermal equilibrium for hot and dense fire ball 
produced in ultra relativistic nuclear collisions, we analyze the evolution 
of the fire ball based on the (3+1)-dimensional
hydrodynamical model. The hydrodynamical model for Quark-Gluon Plasma 
(QGP) fluid has already been discussed in many papers since 
Bjorken first introduced the simple scaling model based on (1+1)-dimensional 
expansion picture \cite{Hydro}. 
For simplicity, cylindrical symmetry is assumed in usual hydrodynamical 
\cite{Ishii,Sollfrank,Hung,Schlei} 
analysis, but this assumption disables us from discussing 
the anisotropic collective flow.
In this paper, in order to investigate collective flow not only 
in the central collisions but also in the non-central ones 
we numerically solve the (3+1)-dimensional relativistic 
hydrodynamical equation coupled with the baryon number 
conservation law.
       
\section{The relativistic hydrodynamical model}
The relativistic  hydrodynamical equation for perfect fluid is 
given as
\begin{equation}
\partial_{\mu}T^{\mu\nu} = 0,
\label{eqn:hydro}
\end{equation}
where $T^{\mu\nu}$ is energy momentum tensor,
\begin{equation}
T^{\mu\nu} = \epsilon U^{\mu}U^{\nu}-P(g^{\mu\nu}-U^{\mu}U^{\nu}).
\end{equation}
Here, $\epsilon$ is energy density, $P$ is pressure,  metric tensor 
is   
$g^{\mu \nu } =
 {\rm diag.} (1,-1,-1,-1)$ and local velocity is $U^{\mu}=
(1,v_{x},v_{y},v_{z})/ \gamma$ respectively.  
In order to take account of the finite baryon number density, 
we must solve the baryon number conservation law,    
\begin{equation}
\partial_{\nu} \{ n_{B}(T,\mu)U^{\nu} \} = 0,
\label{eqn:baryon}
\end{equation}
also, where $n_{B}(T, \mu)$ being baryon number density.
Through the time like component of Eq.(\ref{eqn:hydro}), 
$U_{\nu} \partial _{\mu} T^{\mu \nu}= 0$, Eq.(\ref{eqn:baryon}) and  
thermodynamical relation, $\epsilon +P=TS+\mu n_{B}$, 
we can obtain the conservation law of entropy density current,  
$S^{\mu} = SU^{\mu}$,
\begin{equation}
\partial_{\mu}S^{\mu} = 0.
\label{eqn:entropy}
\end{equation}
Our numerical algorithm solving the hydrodynamical equation 
is based on the entropy 
conservation law Eq.(\ref{eqn:entropy}).

In order to solve the hydrodynamical equation, the equation of state 
is needed.
Though we consider the QGP gas and the hadron gas(excluded volume 
model \cite{Rischke-E})
 for the realistic model equation of states, 
in this paper we adopt the QGP gas for numerical simplicity.
The QGP gas model of massless $N_{f}$ flavor quarks is given by, 
\begin{equation}
P = \frac{32+21N_{f}}{180}\pi^{2}T^{4}
+ \frac{N_{f}}{2}\mu_{q}^{2}T^{2} 
+\frac{N_{f}}{4\pi^{2}} \mu _{q}^{4},
\end{equation}
where the number of flavor being $N_{f}$ = 3 and 
chemical potential for quarks being $\mu_{q} = \mu /3$.  

\section{The numerical calculation}
We solve the (3+1)-dimensional  hydrodynamical equation without 
symmetrical conditions by using an algorithm in which  
lattice points of volume element is moved along local velocity 
and the entropy conservation law Eq.(\ref{eqn:entropy}) 
is adopted explicitly. 
D. H. Rischke et al. discuss relativistic hydrodynamics 
in (3+1)-dimensional situation and collective behavior 
by using Eulerian hydrodynamics \cite{Rischke}.   
Our numerical calculation is explained briefly as follows:

In the first step the coordinates 
$x^{m} = X^{m}(t,i,j,k)$ $(m=1,2,3)$ of lattice points  
 at time   $t+\Delta t $  are replaced by  
\begin{equation}
X^{m}(t+\Delta t,i,j,k)=X^{m}(t,i,j,k) + \frac{U^{m}(t,i,j,k)}
{U^{t}(t,i,j,k)}\Delta t
\label{eqn:coord}
\end{equation}
In the determination of lattice points in Eq.(\ref{eqn:coord}), 
the coordinates move in parallel with 
$n_{B}U^{\mu}$, $SU^{\mu}$.

In the next step the local velocity is determined by, 
\begin{eqnarray}
v^{m}(t+\Delta t,i,j,k)  
 & = &  v^{m}(t,i,j,k) + 
  \partial_{t} v^{t}(i,j,k,t) \Delta t  \\
&  &  +\sum _{m=1} ^{3} \partial_{m}v^{m}(i,j,k,t)
 (X^{m}(t+\Delta t,i,j,k) - X^{m}(t,i,j,k)) \nonumber
\end{eqnarray}
where $\partial_{\mu}v^{\mu}$ obtained from   
Eq.(\ref{eqn:hydro}), Eq.(\ref{eqn:baryon}) is used. 

In the final step the temperature and chemical potential 
of lattice points is calculated by using Eq.(\ref{eqn:baryon}), 
Eq.(\ref{eqn:entropy}).

\section{Comparison with Bjorken's solution}
Comparing our numerical solutions with Bjorken's scaling solution  
$v_{z} = z/t$,  
we can easily evaluate the effect of the transverse flow. 
Based on Bjorken's scaling solution and Eq.({\ref{eqn:entropy}) entropy
density is  given as,   
\begin{equation}
S(\tau) = S({\tau_{0}})\frac{\tau_{0}}{\tau},
\label{eqn:entro}
\end{equation}
where proper time $\tau$, $\tau = \sqrt{t^{2}-z^{2}}$.
In order to make comparison clear, in this section 
the velocity of our model in the longitudinal 
direction is fixed to the Bjorken's scaling solution.
We put the initial temperature distribution and chemical potential 
distribution respectively as follows:
\begin{equation}
T(t_{0} ,x,y,z) = {T_{0}} \exp \left \{
- \frac{(x-{x_{0}})^{2}} {3 \cdot 2 {\sigma _{x}}^{2}} 
\theta (x - {x_{0}}) 
- \frac{(y-{y_{0}})^{2}}{3 \cdot 2 {\sigma _{y}}^{2}}
\theta (y-{y_{0}})
- \frac{(z-{z_{0}})^{2}}{3 \cdot 2 {\sigma _{z}}^{2}}
\theta (z-{z_{0}})
\right \},
\end{equation}
\hspace{0.85cm}
$\displaystyle
\mu(t_{0},x,y,z) = {\mu _{0}}
\left \{
\exp \left [
-\frac{(z-{ z_{B}})^{2}}{2 {\sigma_{B}}^{2}}
\right ]
+ \exp \left [
-\frac{(z+{z_{B}})^{2}}{2 {\sigma _{B}}^{2}}
\right ]
\right \}
\nonumber
$
\begin{equation}
\times \exp \left \{
-\frac{(x-{x_{0}})^{2}}{ 3 \cdot 2 {\sigma _{x}} ^{2}}
\theta (x -{x_{0}})
-\frac{(y-{y_{0}})^{2}}{3 \cdot 2 {\sigma _{y}} ^{2}}
\theta (y - {y_{0}})
\right \},
\end{equation}
where $x_{0}=y_{0}=z_{0}=1.0$ fm, $z_{B}=0.7$ fm, $\sigma _{x}=
\sigma_{y}=\sigma_{z}=1.0$ fm, $\sigma_{B}=0.7$ fm, $T_{0}=200$ MeV, 
$\mu_{0}=210$ MeV and the initial transverse velocity   
is set to 0.  
We focus to the volume elements at $(x,y,z)=(0,0,0)$ for 
the comparison of  
our numerical solution with Bjorken's scaling solution.
\begin{figure}
\begin{center}
\begin{minipage}{.4\linewidth}
\includegraphics[width=0.9\linewidth]{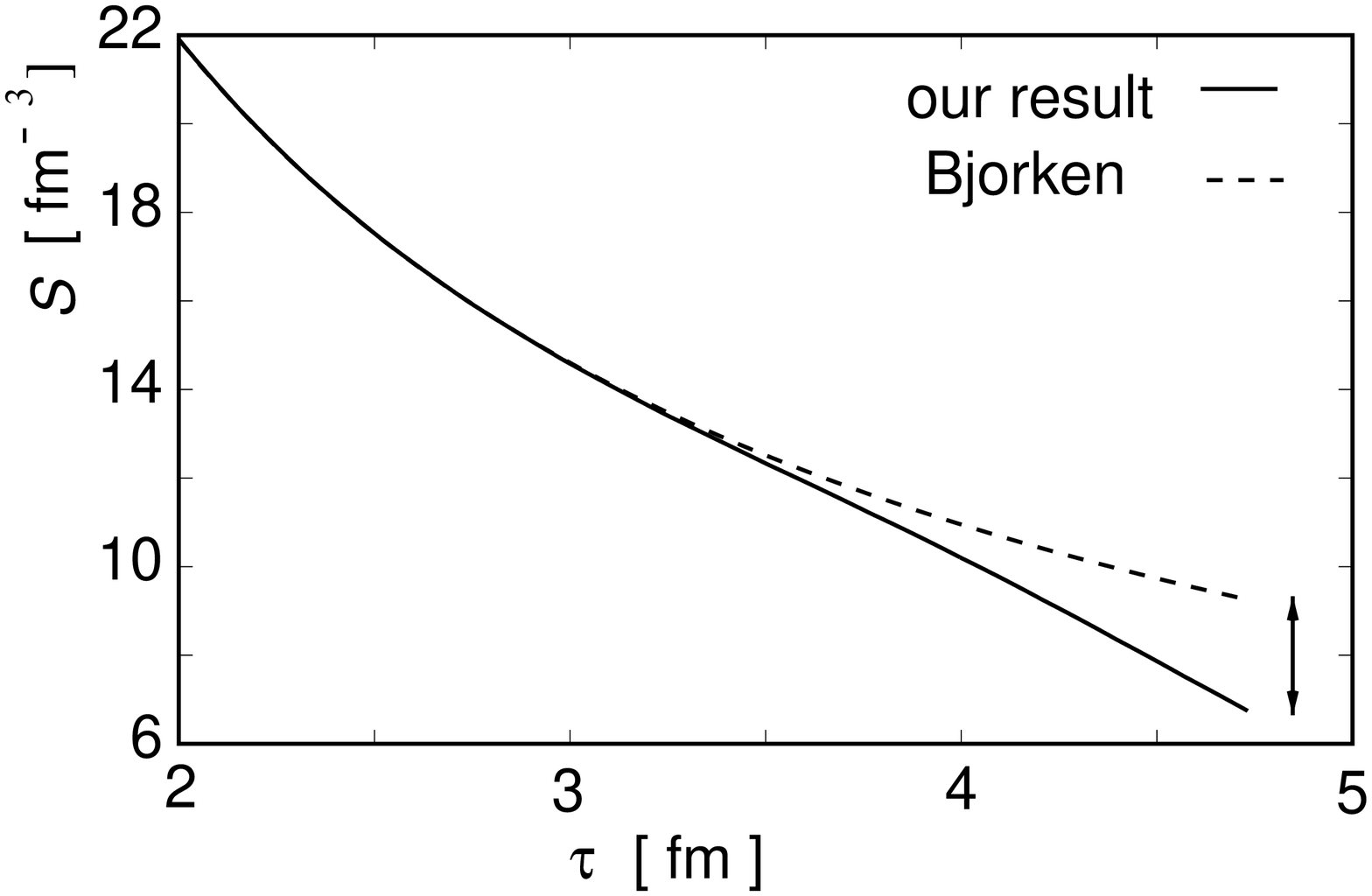}
\caption{the comparison with Bjorken's scaling solution}
\end{minipage}
\hspace{0.4cm}
\begin{minipage}{.4\linewidth}
\vspace{0.6cm}
\includegraphics[width=0.9\linewidth]{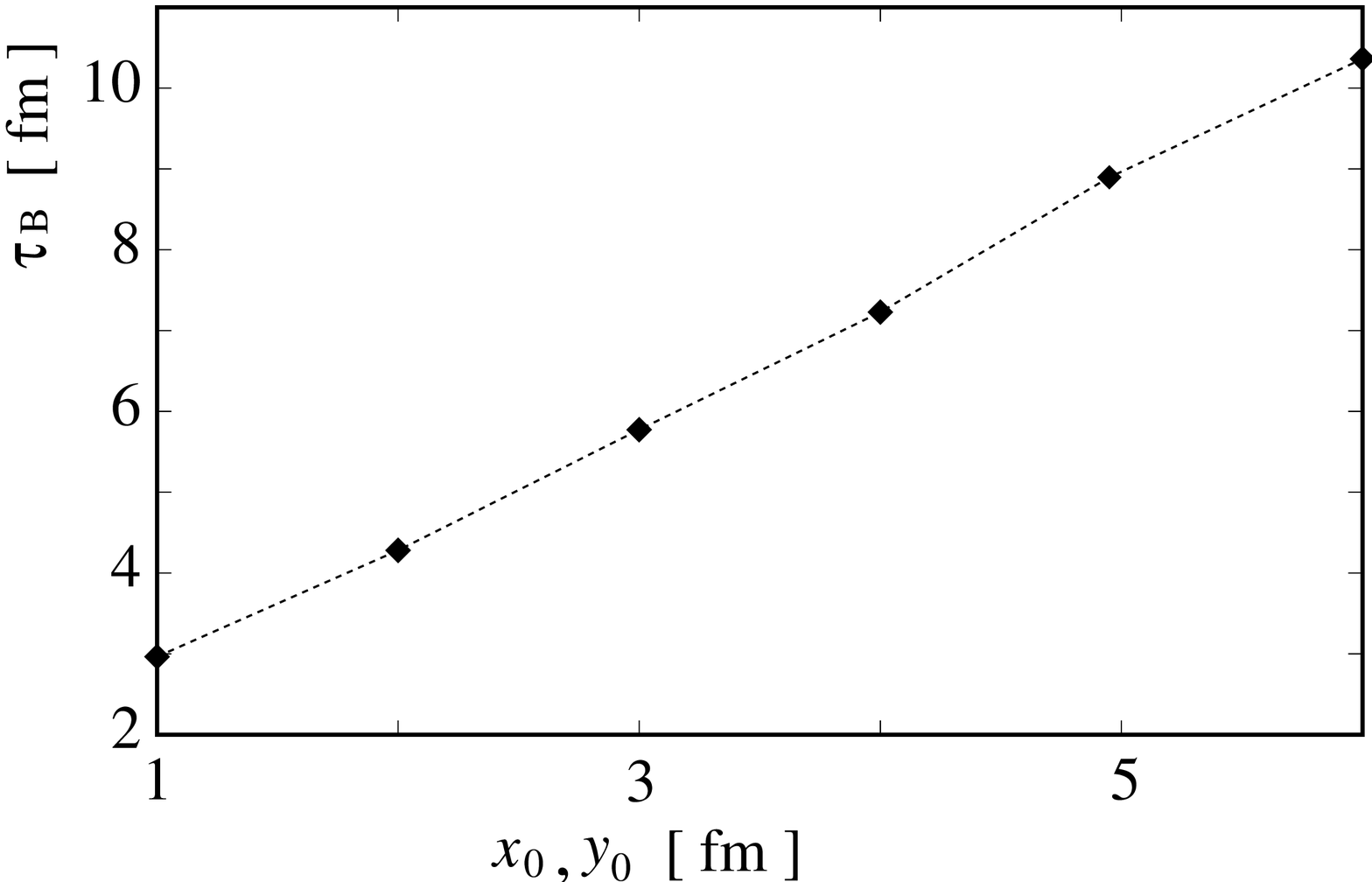}
\caption{the difference between our solution and Bjorken's 
scaling solution by changing $x_{0}$, $y_{0}$ from 1.0 fm to 6.0 fm }
\end{minipage}
\end{center}
\end{figure}
Figure 1 shows that our numerical calculation is 
coincident with Bjorken's scaling solution up to $\tau=3.0$ fm. 
After $\tau=3.0$ fm the difference between them  
increases with the proper time because of the transverse flow.
We define a characteristic time $\tau_{B}$ 
at which instance the difference between two models becomes 
larger than  0.1 \% . 
Figure 2 indicates that $\tau_{B}$ is almost proportional to the
 initial  $x_{0}$, 
$y_{0}$ during  1.0 fm to 6.0 fm,  and Bjorken's 
scaling solution seems to be a proper  solution for a large 
system.   

\section{Non cylindrical initial conditions}

In this section we investigate the space-time evolution of  
the hydrodynamical flow with non cylindrical initial  
temperature distribution and chemical distribution. Other 
parameters are put as the previous section.
\begin{figure}
\begin{center}
\begin{minipage}{0.4 \linewidth}
\includegraphics[width=\linewidth]{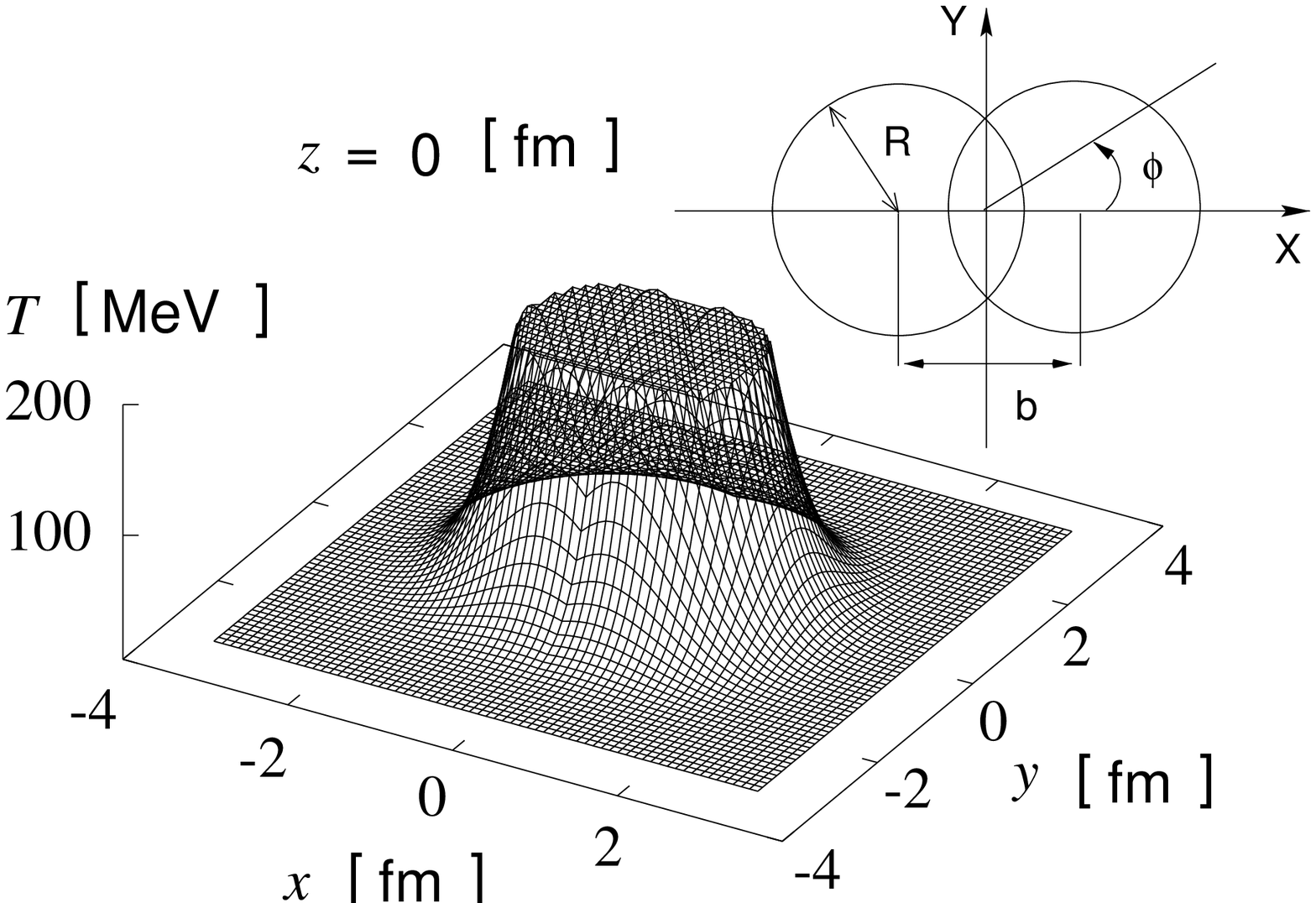}
\caption{the non cylindrical initial condition of temperature
 distribution at $z=0$ fm}
\end{minipage}
\hspace{0.4cm}
\begin{minipage}{0.4\linewidth}
\includegraphics[width=0.9\linewidth]{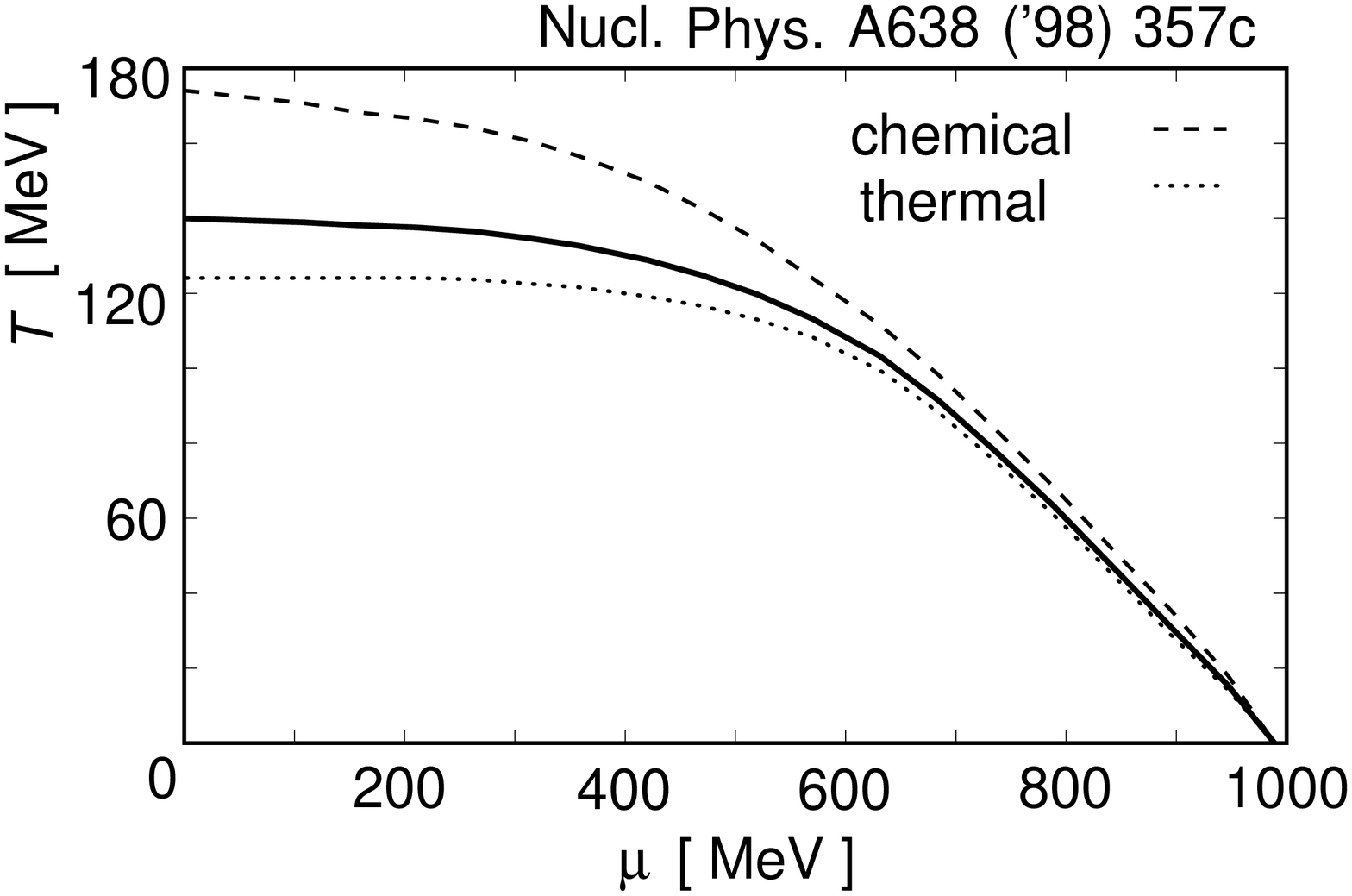}
\caption{The solid line stands for freeze-out which we assume 
from chemical freeze-out (the dashed line) and
 thermal freeze-out (the dotted line). }
\end{minipage}
\end{center}
\end{figure}
Figure 3 shows the initial temperature distribution in $b=0.8$ fm and 
$R=1.0$ fm. 
We evaluate the effect of the flow in particle 
distributions by giving initial conditions like this, 
though this condition is not realistic to analyze 
the experimental data.  
We use Cooper-Frye formula \cite{Cooper-Frye} 
for particle emission from hadronic fluid, 
\begin{equation}
E\frac{dN}{d^{3}P}=\frac{g_{h}}{(2\pi)^{3}} 
\int_{\sigma} d\sigma_{\mu} P^{\mu}
\frac{1}{\exp [(P_{\nu}U^{\nu} - \mu)/T_{\rm f}] \pm 1 }
\end{equation}
 for evaluating one-particle 
distributions. 
We assume that hadronization process occurs when the temperature 
and chemical potential in the volume elements cross the boundary
(the solid line) in fig.4.  
The solid line in fig.4 is so designed that the freeze-out temperature 
becomes 140 MeV at vanishing chemical potential,   
based on chemical freeze-out and thermal 
freeze-out model in ref. \cite{freeze-out}.
Several calculations are made for different initial conditions.
Figure 5 displays the space-time evolution 
of the flow in $b=0.8$ fm.
\begin{figure}
\begin{minipage}{.31 \linewidth}
\begin{center}
 $t=2.5$ fm
\end{center}
\includegraphics[trim=1cm 1.5cm 0.3cm 6.5cm,clip ,width=1.\linewidth]
{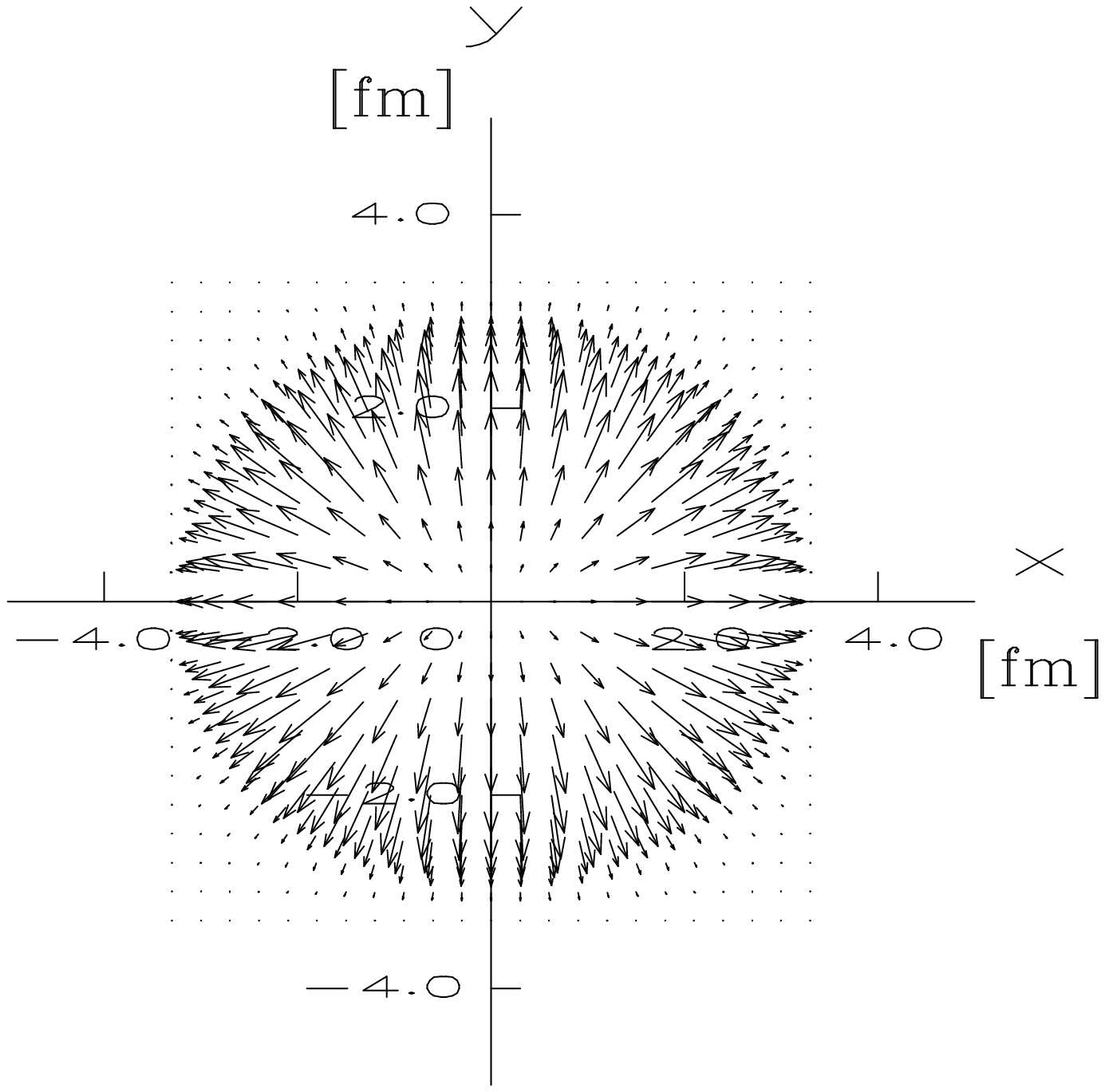}
\end{minipage}
\begin{minipage}{.31 \linewidth}
\begin{center}
$t=3.0$ fm
\end{center}
\includegraphics[trim=1cm 1.5cm 0.3cm 6.5cm,clip ,width=1.\linewidth]
{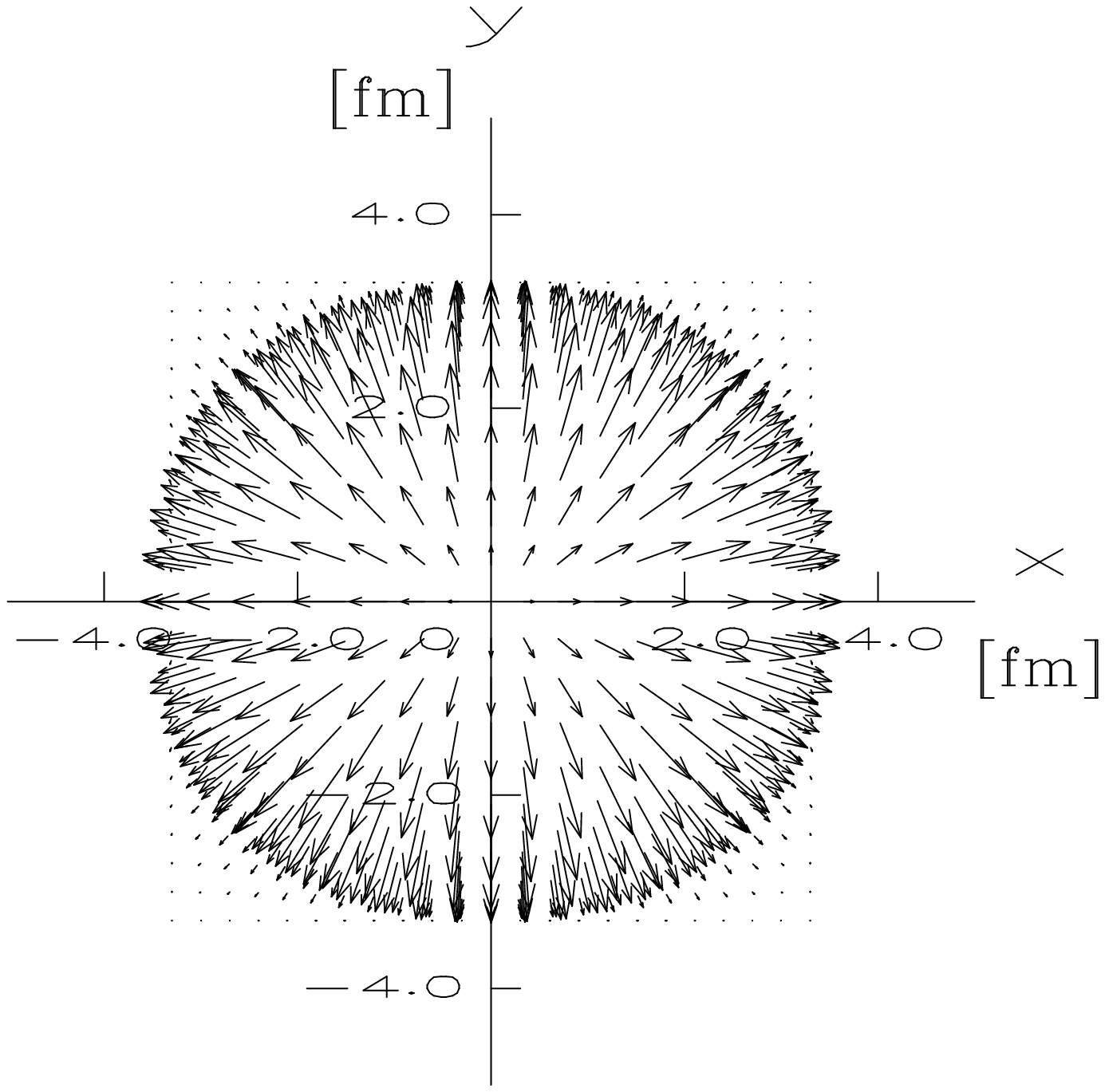}
\end{minipage}
\begin{minipage}{.31 \linewidth}
\begin{center}
$t=3.4$ fm
\end{center}
\includegraphics[trim=1cm 1.5cm 0.3cm 6.5cm,clip ,width=1.\linewidth]
{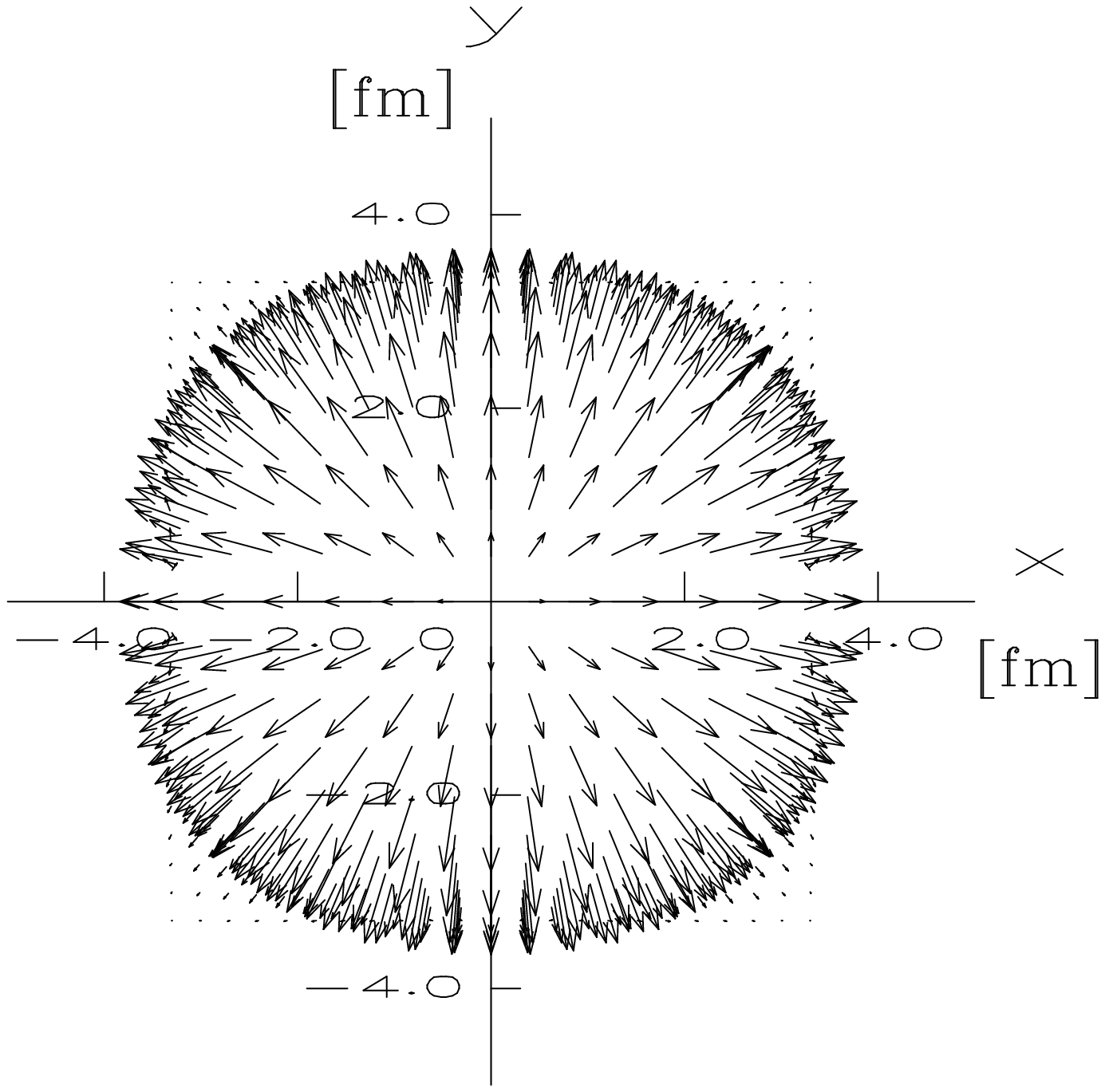}
\end{minipage}
\caption{the space-time evolution of the flow at $z=0$ fm ($b=0.8$ fm)}
\end{figure}
Figures 6 and 7 show the azimuthal fluctuation of particle number 
and $P_{\rm T}$ distribution which are caused by non-cylindrical 
properties of transverse expansion.  
Figure 6 indicates that the variation in the azimuthal distribution   
increase as separation of two initial blobs increases. 
Figure 7 indicates the influence of the flow 
increase with $P_{\rm T}$. 
The yield at $\phi = 90^{\circ}$, 270$^{\circ}$ is large,  
because freeze-out hypersurface is large in these directions 
as fig.5 shows.  Furthermore the transverse momentum distribution 
at $\phi = 90^{\circ}$, $270^{\circ}$ is flatter,  
since the flow is pushed out at $\phi=90^{\circ}$, $270^{\circ}$. 
\begin{figure}
\begin{center}
\begin{minipage}{0.4\linewidth}
\includegraphics[width=0.9\linewidth]{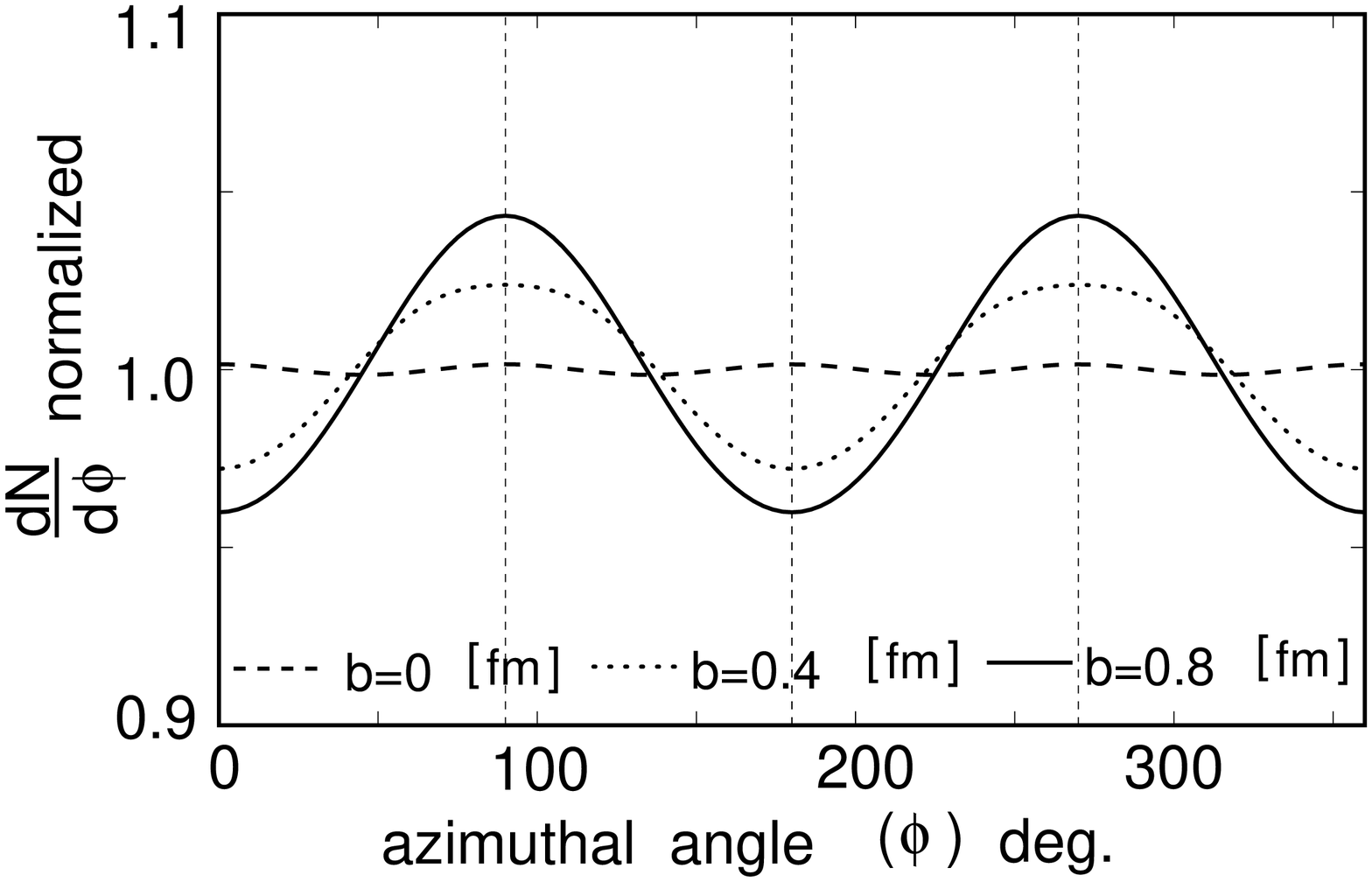}
\caption{the azimuthal distribution in changing $b$}
\end{minipage}
\hspace{0.4cm}
\begin{minipage}{0.4\linewidth}
\includegraphics[width=\linewidth]{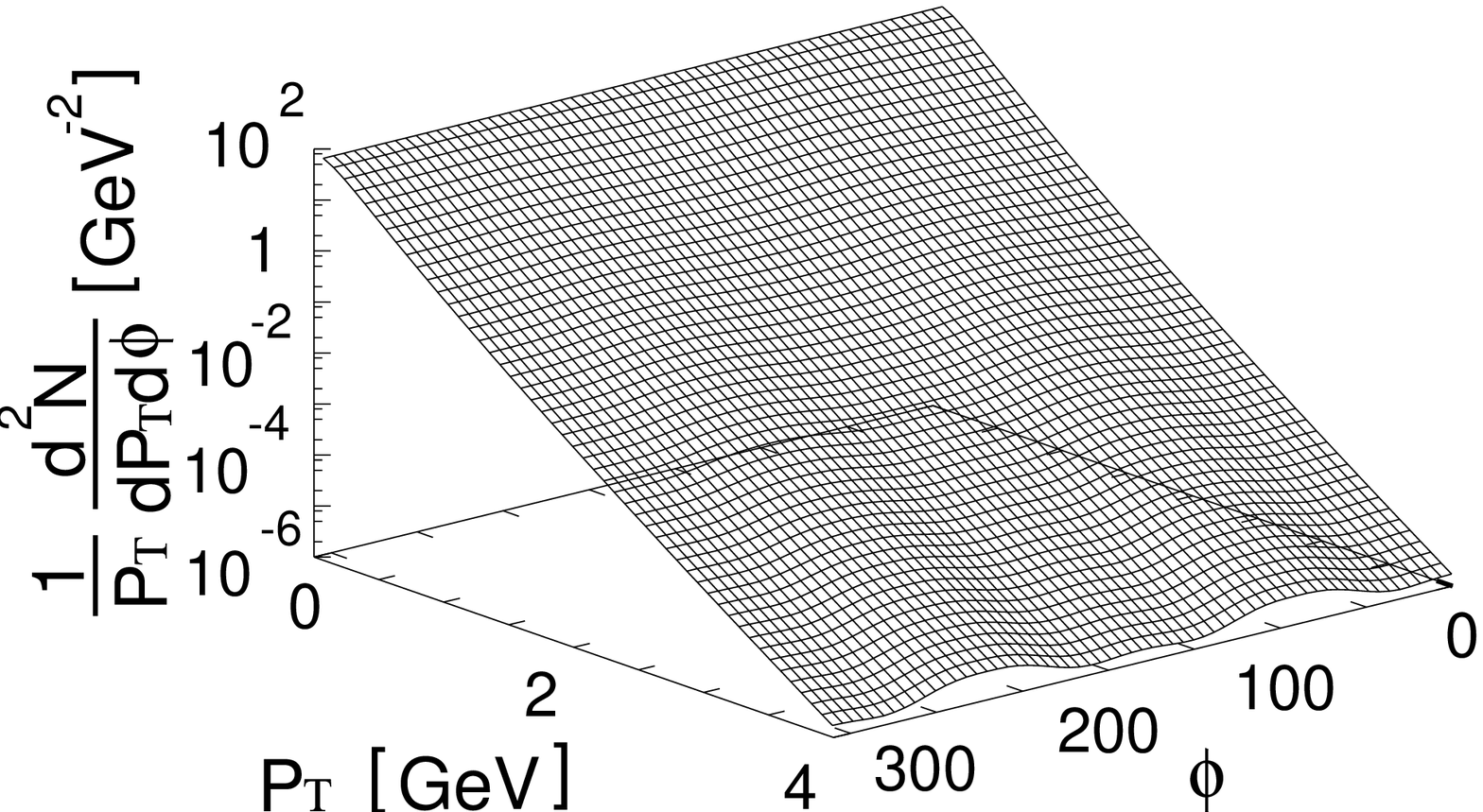}
\caption{the dependence of the transverse momentum distribution on 
the flow ($b=0.8$ fm) }
\end{minipage}
\end{center}
\end{figure}

\section{Summary}
We solved (3+1)-dimensional relativistic hydrodynamical 
equation without cylindrical symmetry conditions 
by Lagrangian hydrodynamics.
We discussed the effect of the transverse flow and  
confirmed that Bjorken's scaling solution is a proper solution 
in a large system   
by making a comparison with numerical calculation. 
The effect of the flow to the particle distributions was 
also investigated.
The influence of flow is large at $\phi=90^{\circ}$, $270^{\circ}$. 
This is because the freeze-out hypersurface is large at
$\phi=90^{\circ}$, $270^{\circ}$ and the flow is pushed out 
in the direction of $\phi=90^{\circ}$, $270^{\circ}$. 
 
We need to use more realistic initial conditions in the  
analysis of the experimental data. We plan to adopt the output 
from the event generator as initial conditions and to use 
the equation of states including phase transition from the 
QGP phase to the hadron phase. 
Investigating the  collective 
flow in experimental data is our next task.

\end{document}